# Spin- and Symmetry-Filtering Combined Tunnel Magnetoresistance through Epitaxial MgO/EuS Tunnel Barriers

Zhiwei Gao, Yihang Yang, Fen Liu, Qian Xue, Guo-Xing Miao

1. Electrical and computer engineering, University of Waterloo, Waterloo, ON N2L 3G1, Canada

2. Institute for Quantum Computing, Waterloo, ON N2L 3G1, Canada

**Abstract**

We created epitaxial magnetic tunnel junctions of FeCo/MgO/EuS on MgO buffered Si (100). Tunnel magnetoresistance reached up to 64% at 4.2 K. An unexpected fast drop of magnetoresistance was recorded for MgO thickness above 1 nm, which is attributed to the forced nonspecular conductance across the EuS conduction band minimum located at the X point, rather than the desired $\Delta_1$ conductance centered around the Γ point.

**Introduction**

Due to the spin selectivity in the quantum mechanical tunneling processes, a magnetic tunnel junction (MTJ) can show a large change in its conductance when its spin configuration changes, namely, the tunnel magnetoresistance (TMR) effect [1,2]. This effect can readily convert useful spin information into detectable electrical signals, therefore MTJs are widely used as highly sensitive magnetic readheads in hard disk drives (HDD), and as nonvolatile data storage units in magnetic random access memories (MRAM). Since M. Jullière first discovered the TMR effect in 1975 [3], it was clear that highly spin-polarized electrodes are desired to improve the TMR response. For decades, the researchers are following this path and better and better performance was achieved in amorphous/polycrystalline materials [4,5,6]. In 2001, Butler and Mathon [7,8] realized that in certain epitaxial systems, such as body-centered-cubic (bcc) Fe coupled with face-centered-cubic (fcc) MgO, the electrons' Bloch wave symmetry can be preserved in the tunneling process, which leads to extremely large TMR [9,10] if the dominating symmetry (such as $\Delta_1$) happens to be fully polarized as well (thus termed symmetry-filtering [11]). An alternative approach to functionalize the tunnel barrier is through spin-filtering [12], where a magnetic tunnel barrier can spontaneously generate spin selectivity and TMR effect without needing magnetic electrodes [13,14]. In this article, we aim to combine the symmetry-filtering effect and the spin-filtering effects to create a hybrid magnetic tunnel junction and investigate its transport performance. In order to have a decent matching, we select epitaxial face-centered-cubic (fcc) MgO coupled with body-centered-cubic (bcc) FeCo as half of the junction. We then pick a suitable spin-filter material - EuS, which shares the same rock-salt structure as MgO, to form the composite barrier. The hybrid junctions showed TMR up to 64% at 4 K.

**Experiment**

The Magnetic Tunnel Junctions were grown on Si (100) wafers. The wafers were cleaned with solvent ultrasonic bath, followed with 1% buffered HF etching before loading into the deposition system (ATC ORION dual chamber deposition system from AJA International). The sample stacks were deposited with high vacuum ebeam depositions, with the system pressure better than $1\times10^{-8}$ Torr during the whole growth process. To promote epitaxy, the wafers were first buffered with 10 nm MgO deposited at 300°C. Then the epitaxial 5.5 nm $Fe_{50}Co_{50}$ / $x$ nm MgO / 3 nm EuS / 10 nm Ti / 3 nm TiN / 3 nm $Al_2O_3$ magnetic tunnel junctions were deposited at slightly above room temperature (100°C) for the best epitaxy correlations. According to the AFM (Atomic Force Microscope) measurement results, we chose 0.08 Å/s to be the growth rate of each layer to ensure small roughness and better quality in our junctions. The samples were *in-situ* patterned using shadow masks with the active junction areas being 30×30 µm$^2$. All the samples are sealed in dry $N_2$ environment till the transport measurement. The transport features of these junctions were done with 2-terminal method by sweeping magnetic field in the range of -300 G and 300 G. Due to the high impedance of the double barrier junctions, 2-terminal method does not generate much error in the results. The samples were submerged in liquid helium bath in the transport measurement.

**Results and Discussion**

The epitaxy of MgO on Si(100) has been well demonstrated previously with 300°C deposition [15]. MgO cells sit straight on top of Si cells with a 4-on-3 alignment, and a lattice mismatch of 3.4%. We chose 10 nm MgO in this experiment to fully relax the strain, and also to eliminate the leakage conductance through Si substrate in measuring high impedance junctions. We next chose $Fe_{50}Co_{50}$ as the bottom electrode instead of Fe, because it leads to higher spin polarization when coupled with MgO [16]. But when we tried to deposit FeCo onto the MgO buffer layer at room temperature, we found weak extra FeCo (110) peaks under XRD. After trying to grow FeCo at several higher temperatures, we found that the best condition is at approximately 100°C to get only the desired (100) orientation. We try not to increase the temperature beyond necessary as a balance to maintain smooth interfaces. Under this condition, we found only the FeCo (200) peak as one can see in Fig.1, top panel.

The bulk lattice parameters of MgO and FeCo are 4.212Å and 2.857Å, they have a ratio of approximately $\sqrt{2}$, which indicates that FeCo cells are 45° rotated in plane relative to both Si and MgO cells in order to form the best lattice matching. The next

MgO layer on FeCo is then 45° rotated back. EuS has a lattice parameter of 5.968Å and lattice mismatch of -5.5% with MgO. Had EuS be in-plane rotated by 45° with respect to MgO, the lattice mismatch would be only 0.2%; the similarity of the two rock-salt ionic crystals clearly prevents such rotation to happen. We used off-axis Φ scans on the FeCo {110} and EuS {220} reflections to verify the above stated relative orientations (middle and bottom panels in Figure 1). The sample was tilted by 45°, then spin around its norm (Φ scan) to yield the group of {110} reflections to the detector. And further θ-2θ scans were performed at the individual Φ peak positions to verify the labelled peak assignments. Setting the θ-2θ angles towards the desired reflections will reveal only those reflections (ideally); and a cubic, (100) oriented epitaxial material will show the {110} reflections four times under 360° Φ rotation. On both the FeCo{110} and EuS{220} scans, the extremely strong and sharp peaks are from Si{220}, which show up nevertheless because they are too strong coming from the single crystal wafer. From the relative positions of these peaks, we can confirm that FeCo is indeed 45° rotated in-plane relative to EuS and Si. The diffraction angles of FeCo{110} and EuS{220} are broad and close to each other, therefore detected on both Φ scans. The peak broadening suggests that the epitaxy gets worse as the layers progress, probably from mosaic growth and small-angle-grain-boundaries formation due to the large lattice mismatch between layers.

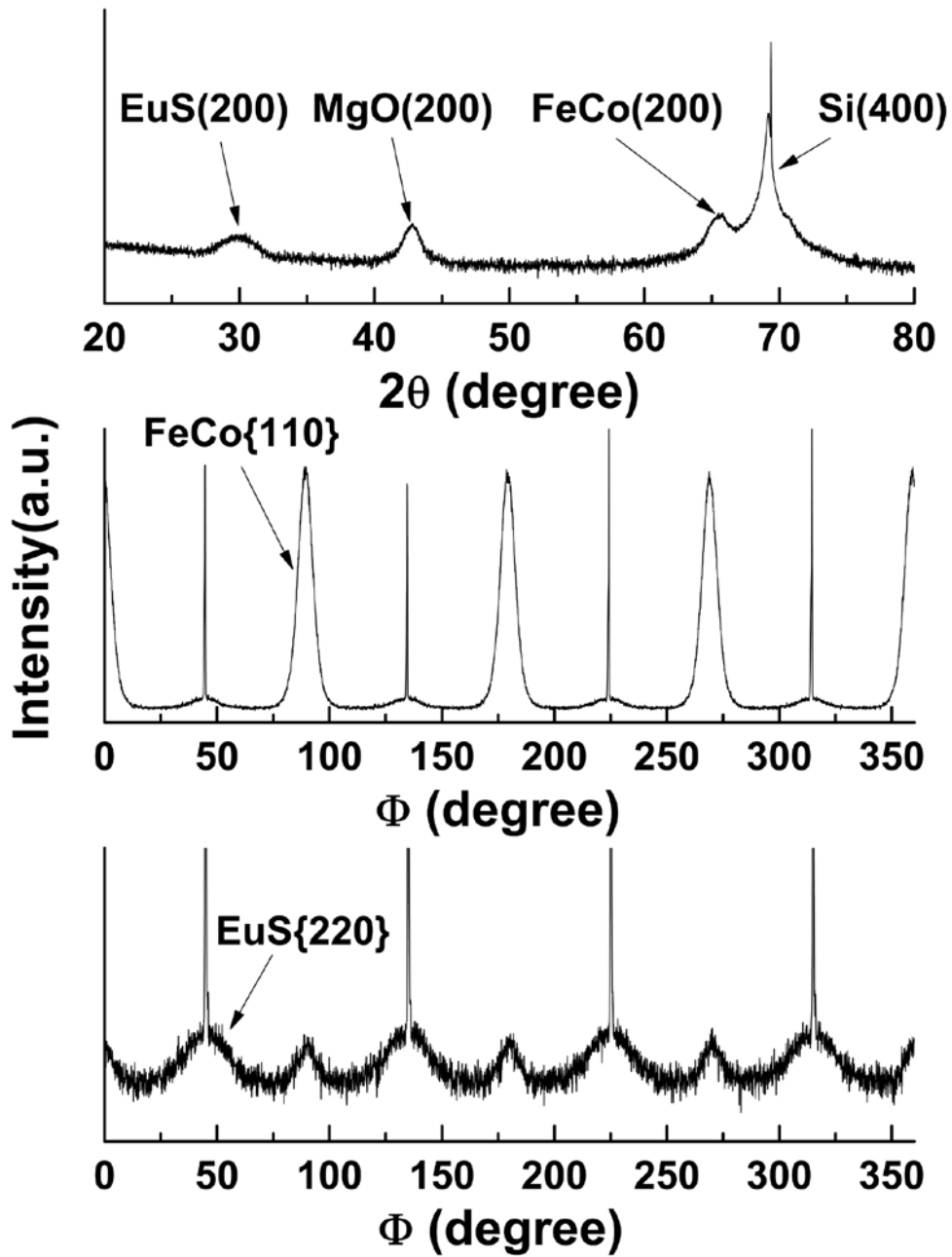

Figure 1: Top panel shows the XRD θ-2θ scan of the magnetic tunnel junction: Si (100) / 10 nm MgO / 5.5 nm FeCo / 3 nm EuS. The bottom two panels are the off-axis Φ

scan targeting FeCo{110} and EuS{220} peaks (2θ set at 44.83° and 42.84°), respectively.

3 nm EuS has been shown to be the optimal thickness choice for generating TMR in EuS based spin-filter tunnel junctions [14,17]. Being too thin will reduce its spin-filtering efficiency, which goes up with thickness; while being too thick will hinder the direct tunneling. In our experiment, we kept the EuS layer thickness at 3 nm, and only vary the MgO thickness. The best TMR we obtained was at 1 nm MgO, with the junction resistance around 30 kΩ at room temperature and over MΩ at low temperatures. This is consistent with EuS's semiconductor nature (electrical band gap ~ 1.6 eV). Figure 2 summaries the dependence of magnetoresistance with respect to bias voltage for these junctions, with the highest recorded TMR being 64%. This value is consistent with what one would expect from the known spin polarizations of 1 nm MgO [9] and 3 nm EuS [14]. All the samples showed fast TMR decrease with increasing voltage, a typical behavior when the barrier is less than perfect. The most striking phenomena we observed on this type of junctions are the strong dependence of TMR on MgO thickness, with all other parameters fixed (Figure 2, upper left inset). The initial increase of TMR with increasing MgO thickness is quite justified, as the symmetry-filtering capability of MgO relies on the faster exponential decay of non-$\Delta_1$ states within the barrier. The enhancement through symmetry-filtering is expected to continue till at least 2 nm, beyond which further increase has limited effect and TMR would saturate [9]. However, we observed a significant drop of TMR beyond 1 nm MgO thickness, which cannot be attributed to saturation of MgO's symmetry-filtering, nor defect mediated hopping. Both of these effects, widely present in conventional MTJs, only show a gradual change of TMR as the barrier gets thicker and thicker. The pronounced thickness dependence in our junctions is clearly originated from other mechanisms.

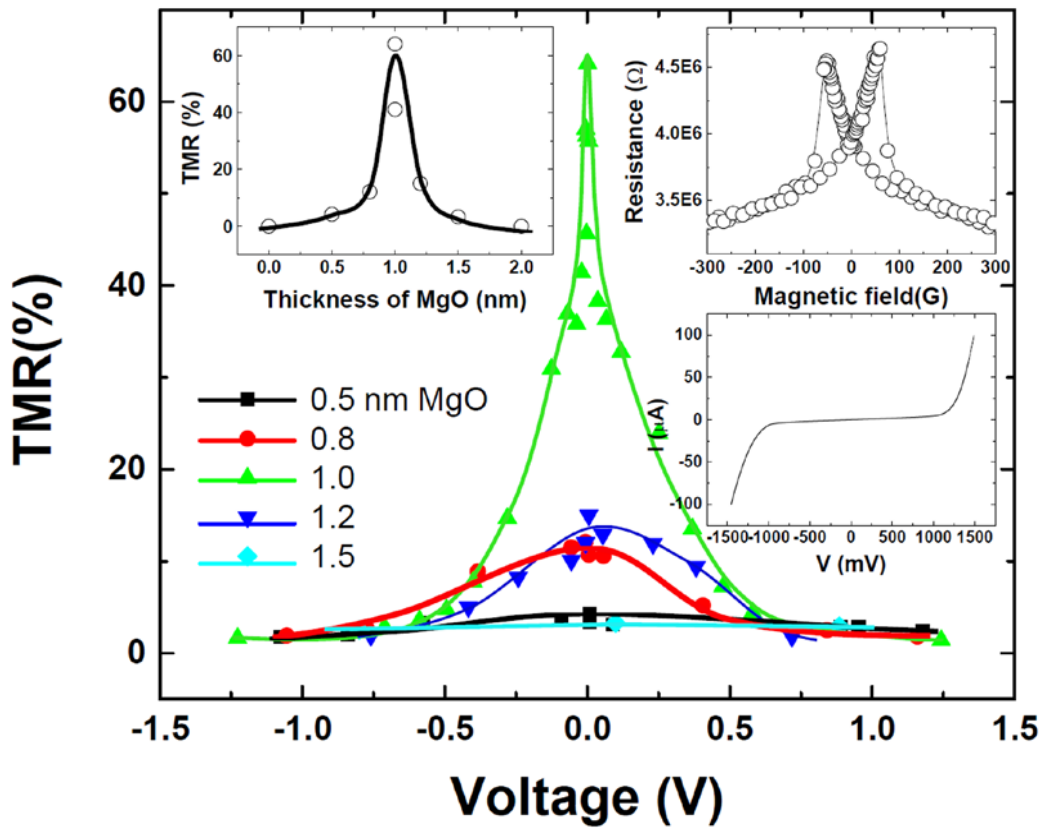

Figure2 TMR response with respect to bias voltages for different thickness of MgO. The upper left inset shows that maximum TMR occurs at around 1nm MgO, further increasing or decreasing the thickness yields smaller TMR. The upper right inset shows a typical magnetoresistance response of such junctions, and the lower right inset a typical IV curve.

For junctions formed with $Al_2O_3$ and EuS [14], a strong TMR enhancement was observed with the onset of Fowler-Nordheim tunneling above 1 V. However, as shown in this study, TMR dropped to practically zero before 1 V when epitaxial MgO and EuS are used. Clearly significant scattering happened in this type of junctions despite the improved crystalline quality. The enhanced scattering can be traced to the band structures of the two types of materials. Due to the similarity of the crystal structures of EuS and MgO (both being ionic rock-salt crystals), we expect the electrons to readily maintain their Bloch wave symmetries in the tunneling process, which should yield high spin transport efficiency. Among the Bloch states relevant to the transport, the $\Delta_1$ states have the slowest decay rate. As MgO grows thicker, $\Delta_1$ states become dominant and they happen to be fully spin-polarized coming from FeCo, therefore TMR generally increases. We make a schematic comparison between the band structures of MgO and that of EuS [18] (Figure 3). MgO has a standard direct gap with *p*-electrons forming the valence bands, and the *s*-electrons the conduction bands. The *spd* electrons with projections symmetric around the Γ point jointly form the evanescent $\Delta_1$ states [11]. As a comparison, it is clear that the Γ point is not the conduction band minimum in EuS. Even though symmetry still permits the $\Delta_1$ states to tunnel through, additional tunneling is permitted at the X point with a much reduced barrier height. As a result, when nonspecular scatterings happen inside the barrier [19], they are granted an easier path out therefore an enhanced (but undesired) conductance. More complications arise because the valence band maximum is formed with 4$f$ bands, which do not match any of the bands coming from FeCo/MgO. From the above discussions, we see that the $\Delta_1$ symmetry tunneling prefers to travel near the Γ point, while the EuS spin filtering prefers electrons going along the X point. As a result, undesired scatterings are forced to happen in the tunneling process. Because of the exponential dependence of tunnel probability on tunnel barrier heights, these conduction channels quickly become dominant and the TMR diminishes. Though this study uses a rare-earth chalcogenide as the spin-filter, such band mismatch is more of a common feature in many popular spin-filter materials, such as the ferrites [20] and perovskites with indirect gaps [21]. When the active spin-filtering bands are displaced from the Γ point, scattering becomes more probable and the combination of these two very efficient spin systems however would lead to reduced TMR instead.

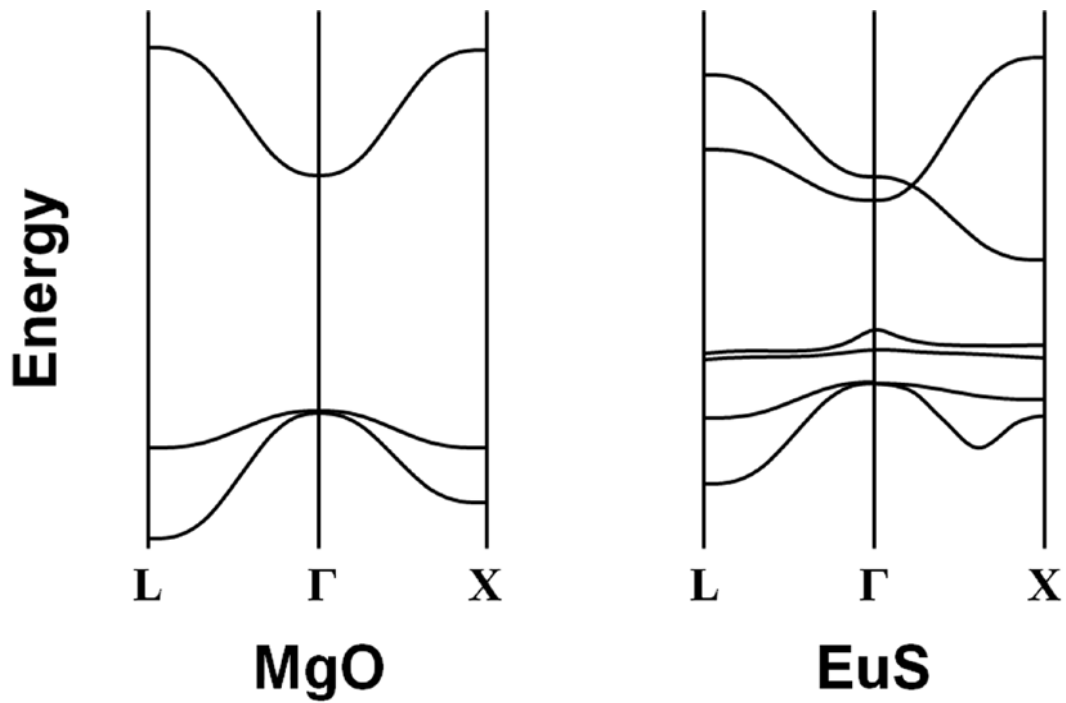

Figure3 Schematics of the band structures of MgO and EuS (exchange splitting not shown, which is roughly a rigid shift). The $\Delta_1$ symmetry tunneling prefers electrons to travel near the Γ point, while the EuS spin filtering prefers electrons going along the X point.

**Conclusion**

We have studied symmetry-filtering and spin-filtering combined magnetic tunnel junctions, FeCo/MgO/EuS/Ti. The best TMR we achieved is 64% at 4K. Remarkably, TMR starts to drop quickly as the MgO thickness goes beyond 1 nm, whereas its symmetry-filtering is expected to become stronger. This behavior is attributed to the displaced conduction band minimum of the spin-filters, and the consequently increased nonspecular transport in the tunneling process.

**Acknowledgement**

This work was supported by NSERC Discovery grant RGPIN 418415-2012.